\documentclass[sigconf,10pt]{acmart}

\usepackage{booktabs} % For formal tables
 \usepackage{enumitem}
\usepackage{balance}

\setcopyright{none} 
\renewcommand\footnotetextcopyrightpermission[1]{} % removes footnote with conference information in first column
\begin{document}
\title{Automatic Model Parallelism for Deep Neural Networks with Compiler and Hardware Support}
%\titlenote{Produces the permission block, and
%  copyright information}
%\subtitle{Extended Abstract}
%\subtitlenote{The full version of the author's guide is available as
%  \texttt{acmart.pdf} document}

\author{Sanket Tavarageri}
\affiliation{%
  \institution{Intel Labs}
}
\email{sanket.tavarageri@intel.com}

\author{Srinivas Sridharan}
\affiliation{%
  \institution{Intel Labs}
}
\email{srinivas.sridharan@intel.com}

\author{Bharat Kaul}
\affiliation{%
  \institution{Intel Labs}
    }
\email{bharat.kaul@intel.com}


\begin{abstract}
The deep neural networks (DNNs) have been enormously successful in tasks  that were hitherto in the human-only realm such as image recognition, and language translation. Owing to their success the DNNs are being explored for use in ever more sophisticated tasks. One of the ways that the DNNs are made to scale for the complex undertakings is by increasing their size -- deeper and wider networks can model well the additional complexity. Such large models are trained using model parallelism on multiple compute devices such as multi-GPUs and multi-node systems.

In this paper, we develop a compiler-driven approach to achieve model parallelism. We model the computation and communication costs of a dataflow graph that embodies the neural network training process and then, partition the graph using heuristics in such a manner that the communication between compute devices is minimal and we have a good load balance. The hardware scheduling assistants  are proposed to assist the compiler in fine tuning the distribution of work at runtime. 
\end{abstract}

%\begin{CCSXML}
%<ccs2012>
%  <concept_id>10010520.10010553.10010562</concept_id>
%  <concept_desc>Computer systems organization~Embedded systems</concept_desc>
%  <concept_significance>500</concept_significance>
%  <concept_id>10010520.10010575.10010755</concept_id>
%  <concept_desc>Computer systems organization~Redundancy</concept_desc>
%  <concept_significance>300</concept_significance>
%  <concept_id>10010520.10010553.10010554</concept_id>
%  <concept_desc>Computer systems organization~Robotics</concept_desc>
%  <concept_significance>100</concept_significance>
%  <concept_id>10003033.10003083.10003095</concept_id>
%  <concept_desc>Networks~Network reliability</concept_desc>
%  <concept_significance>100</concept_significance>
%</ccs2012>
%\end{CCSXML}

%\ccsdesc[500]{Computer systems organization~Embedded systems}
%\keywords{ACM proceedings, \LaTeX, text tagging}

\maketitle

\section{Introduction}
The deep neural networks (DNNs) as they grow in size necessitate the use of multiple compute devices (e.g., multi-GPUs) for their training. When the neural network model is split across multiple compute devices while training, it is termed model parallelism. Achieving high performance in model parallelism is an important and a difficult problem. We have developed a comprehensive solution to automatically obtain efficient model parallelism through compiler analyses and with the use of novel hardware support.

%\section{Automatic Model Parallelism for Deep Neural Networks with Compiler and Hardware Support}

\begin{figure*}[h!]
\centering
\includegraphics[scale=0.55]{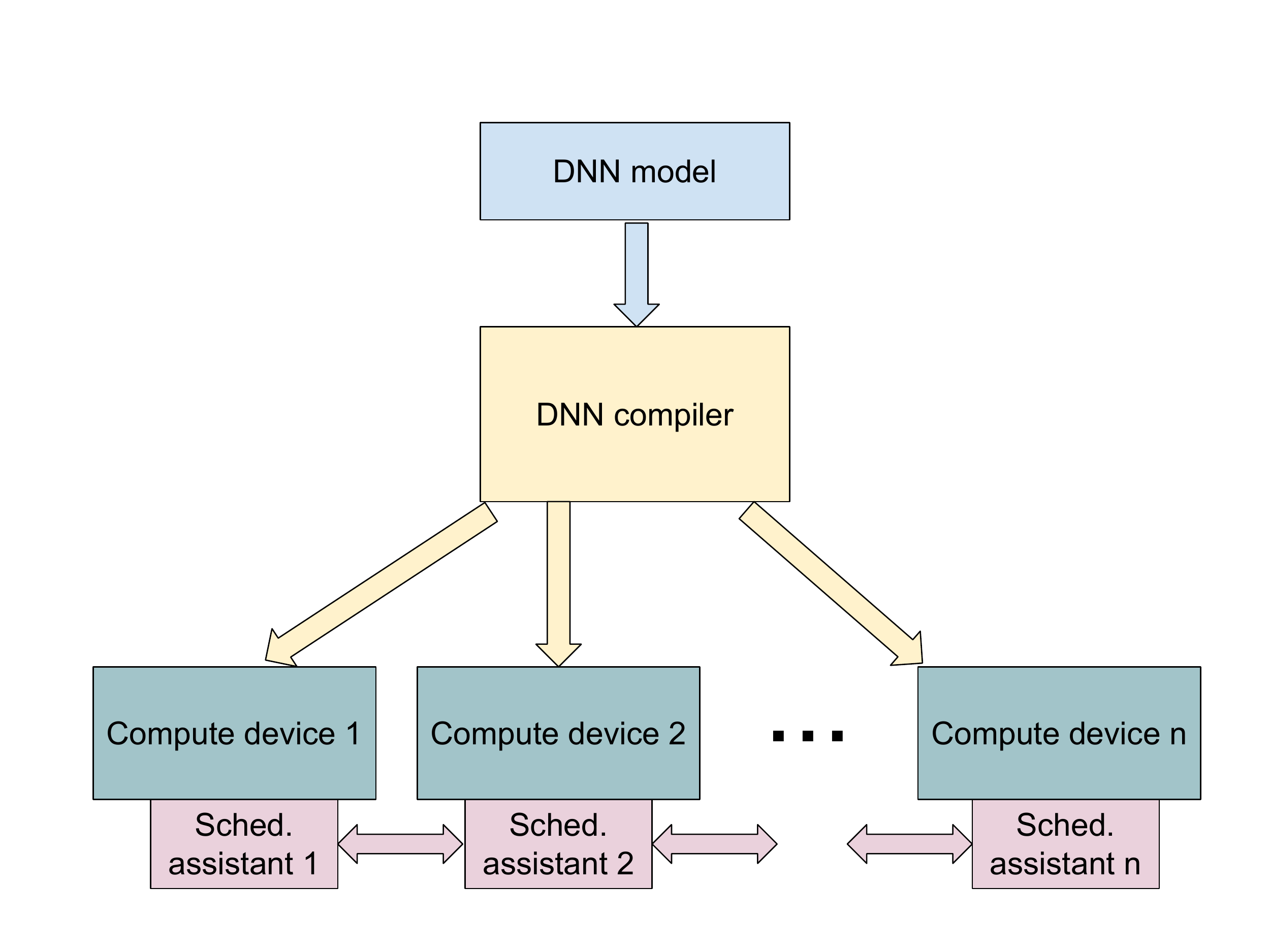}
\caption{The overview of the system}
\label{fig:system}
\end{figure*}

We develop a compiler and hardware scheduling assistant based solution for realizing model parallelism while training deep neural networks (DNNs). Figure \ref{fig:system} shows the overview of the system. The DNN compiler maps the dataflow graph produced by the DNN model to multiple compute devices to efficiently execute the graph.
The hardware scheduling assistants are programmed by the compiler to optimally migrate computations between devices at runtime so that the resources of the system are effectively utilized.

The DNN compiler through analytical cost modeling of computation and communication costs, partitions and maps the neural network to multiple compute devices that achieves minimal communication and optimal load balance. However, to account for impreciseness in analytical cost modeling, and dynamic changes in the execution environment, it enlists the hardware help as follows. The compiler encodes simple rules for hardware scheduling assistants to move around parts of the neural network to dynamically adapt for high performance. 

The techniques presented in this paper will dramatically increase the performance of training of deep learning models on Intel architectures. The hardware and software synergy that this solution will bring about, will be effective in achieving better scaling compared to compiler-only approaches as the system will be able to adapt to changing execution environments and will fine-tune the model parallelism for performance continually.  This hardware, software co-design is superior to other middle-ware based runtime techniques because of encoding of rules in the hardware which will eliminate the runtime overheads.

In Section \ref{section:compiler}, we describe the compiler techniques, and the hardware scheduling assistants are detailed in Section \ref{section:hardware_support}. The related work is discussed in Section \ref{section:related}, while concluding remarks are presented in Section \ref{section:conclusion}.

\section{Compiler directed model parallelism}
\label{section:compiler}
In deep neural network frameworks such as TensorFlow, the computation is represented as a dataflow graph.
The nodes of the graph represent computations while the edges capture the input and output data.
To execute the dataflow graph on multiple devices, such as in a multi-GPU, or a multi-node environment, first we need to identify which parts of the graph will run on what device, and also have to insert communication primitives between any two nodes that are connected by an edge but are now mapped to different devices. 

In this work, we develop an approach to partitioning the dataflow graph. The subgraphs that the partition induces will be executed on different devices. The goals of the partitioning algorithm will be to 1) reduce the volume of data to be communicated between subgraphs, and 2) achieve a good load balance by creating subgraphs of roughly equal size.

The various phases of the overall approach are as follows.
\begin{enumerate}[leftmargin=0pt]
\item \textbf{Selection of computationally expensive and relocatable nodes:} 
We profile the workload to discover computationally expensive nodes.
Further, among the computationally expensive nodes, only the stateless nodes are considered for further analysis (an example of a stateful node would be a variable which is used to save a model's parameters).

\item \textbf{Analytical cost modeling:}
Analytical cost modeling of the dataflow graph is performed which assigns computation, and communication costs to nodes, and edges of the graph respectively.
Here, cost modeling of only the selected computationally expensive nodes is carried out.
The costs thus assigned form the basis of subsequent graph partitioning. 

The dataflow graph $G$ consists of vertices/nodes $V$ and edges $E$: $G = (V, E)$. 
The nodes indicate computations, and the edges encode the data and control dependencies between nodes.
Let there be $k$ devices available: $\mathcal{D}_1, \mathcal{D}_2, \dots \mathcal{D}_k$ with potentially varying computational capabilities.
A node $v_i$ mapped to a device $\mathcal{D}_j$ is assigned a computation cost of $c_{v_i}^{\mathcal{D}_j}$. The cost $c_{v_i}^{\mathcal{D}_j}$ denotes the number of time units it takes to execute $v_i$ on  
$\mathcal{D}_j$. The cost modeling is based on the number of operations an operator entails, and the number of operations that a given device can perform in a unit of time.

The edges of the graph are assigned numerical weights equal to the volume of data that the edges carry. The edges denote data and control dependencies between nodes. A control dependency edge is given the weight 0, while the number of bytes of data an edge carries becomes its weight. In terms of notation, the weight $d_{v_i v_j}$ is assigned to the edge connecting node $v_i$ and node $v_j$.

\item \textbf{Initial partitioning of the dataflow graph:}
We use one of the following strategies to create initial partitions: 1) block partitioning, 2) random partitioning. The heuristic described next subsequently improves the partitions in terms of optimizing criteria, namely communication minimization, and load balance.

\emph{Block partitioning:} If $\mathcal{C}$ is the total computation cost, and $k$ is the number of devices, then the nodes are assigned to devices in such a way that each partition gets nodes worth a total of $\frac{\mathcal{C}}{k}$. To do so, the dataflow graph is topologically sorted, and a list of sorted order of nodes is created. Then, the list is divided up into $k$ partitions in a block fashion so that nodes in each partition have an aggregate cost of $\frac{\mathcal{C}}{k}$. The $k$ partitions are mapped to $k$ devices.

\emph{Random partitioning:} The nodes are randomly assigned to devices.

\item \textbf{Iterative repartitioning}
We adapt the Kerningham and Lin formulation  of the communication cost \cite{kernighan1970efficient} and Karypis and Kumar's greedy refinement approach \cite{karypis1998multilevelk} for the context of dataflow graphs which are directed graphs. 
(The Kerningham and Lin formulation is applicable only to undirected graphs; unlike Karypis and Kumar's approach where load balance is a secondary goal, in our formulation we can consider it to be a primary goal which allows us to completely automatically achieve model parallelism).

The communication cost of a node $n_i$ mapped to device $\mathcal{D}_p$ is calculated as follows. The incoming edges into $n_i$ are considered. Let $I_{n_i}$ be the sum of weights of edges emanating from nodes that are mapped to the same device as $n_i$ and end in $n_i$.
Let $E_{n_i}$ be the sum of weights of edges originating from nodes mapped to a device different from that of $n_i$. The difference between $E_{n_i}$ and $I_{n_i}$ is the communication cost associated with $n_i$.

$$D_{n_i}^{\mathcal{D}_p} =  E_{n_i}^{\mathcal{D}_p} - I_{n_i}^{\mathcal{D}_p}$$

It is observed that if $E_{n_i}^{\mathcal{D}_p}$ is 0 then $D_{n_i}^{\mathcal{D}_p}$ is a negative value assuming $I_{n_i}^{\mathcal{D}_p}$ is non-zero. In this case all of $n_i$'s communication is internal to the device. On the other hand, if $I_{n_i}^{\mathcal{D}_p}$ is 0, then $D_{n_i}^{\mathcal{D}_p}$ is a positive quantity assuming $E_{n_i}^{\mathcal{D}_p}$ is non-zero. In this instance,  all of $n_i$'s communicating partners are located on other devices.

We would like to minimize the sum of $D_{n_i}$s over all nodes as much as possible to achieve minimal communication subject to the constraint that a certain load balance requirement among devices is maintained. We define the load balance constraint as the share of the computational cost of a device being within a threshold of the \emph{ideal share} of the computational cost. That is,

$$ \forall_{\mathcal{D}_i \in \{\mathcal{D}_1, \dots \mathcal{D}_k \}}   \biggm| C_{\mathcal{D}_i} - \frac{\mathcal{C}}{k} \biggm| \leq \epsilon $$

where $\epsilon$ is a parameter.

We move a node $n_i$ from device $\mathcal{D}_q$ to device $\mathcal{D}_r$ with the \emph{minimum} $D_{n_i}$ and if the following condition is met:

$$ \bigg( D_{n_i}^{\mathcal{D}_r} < D_{n_i}^{\mathcal{D}_q} \bigg) \land   \bigg( (C_{\mathcal{D}_r}  +  c_{n_i}^{\mathcal{D}_r}) - \frac{\mathcal{C}}{k}  \leq \epsilon \bigg) \land \bigg( \frac{\mathcal{C}}{k} - (C_{\mathcal{D}_q}  -  c_{n_i}^{\mathcal{D}_q})   \leq \epsilon \bigg)  $$

The first part of the condition makes sure that the communication cost on the new device  $\mathcal{D}_r$ is smaller than the communication cost on the original device  $\mathcal{D}_q$. The second part of the formula states that the computation share of the device $\mathcal{D}_r$ receiving the new node does not exceed the ideal computational share beyond the threshold $\epsilon$. The third part of the conjunction asserts that the computational share of the device $\mathcal{D}_q$ losing the node does not drop below the threshold $\epsilon$ when compared to the ideal share.

In addition to communication minimization, the additional goal is to also improve the load balance of the system, a node $n_i$ is moved from device $\mathcal{D}_q$ to device $\mathcal{D}_r$ if 1) device $\mathcal{D}_q$'s consequent computational share remains above the ideal share, and 2) device $\mathcal{D}_r$'s share remains below the ideal share:

$$ \bigg( (C_{\mathcal{D}_r}  +  c_{n_i}^{\mathcal{D}_r}) < \frac{\mathcal{C}}{k} \bigg) \land \bigg(  (C_{\mathcal{D}_q}  -  c_{n_i}^{\mathcal{D}_q}) > \frac{\mathcal{C}}{k} \bigg)  $$

\section{Hardware Support: Scheduling Assistants}
\label{section:hardware_support}
Owing to the impreciseness of the analytical model, and possible interference of co-located applications, the compiler directed model parallelism may not be able to achieve optimal performance. Therefore, we augment compiler orchestrated model parallelism with dynamic adaptation by hardware scheduling assistants. The scheduling assistants are programmed by the compiler with a set of rules that will dictate how the nodes are migrated among compute devices.

The nodes in the dataflow graph will be annotated with the following tags depending on the bottleneck that the operations in the nodes face:
\begin{itemize}
 \item \emph{\textbf{compute-bound}}
\item \emph{\textbf{memory-bound}}
\item \emph{\textbf{network-bound}}
\end{itemize}

The scheduling assistant observes the compute, memory, and network activity on a device, and migrates the nodes depending on their tags as follows:

\begin{itemize}
 \item When a device $\mathcal{D}_i$'s compute utilization exceeds a certain threshold $\theta$ (say, 95\%), then it selects one of the compute-bound nodes mapped to it and places it in the \emph{compute out-box}. Another device whose compute utilization falls below a certain threshold $\gamma$ (say, 50\%) may acquire the node thus placed in the compute out-box of $\mathcal{D}_i$.
\item Correspondingly, if the compute utilization of $\mathcal{D}_i$ falls below $\gamma$, then $\mathcal{D}_i$ picks a node placed in another device's compute out-box.
\end{itemize}

Similar rules are formed with respect to memory bandwidth utilization, and network utilization. The compiler's designating of nodes as compute-bound, or memory-bound, or network-bound provides the scheduling assistants of the system to swap nodes of the dataflow graph to maximize their collective utilization of various resources.

\end{enumerate}

\section{Related Work}
\label{section:related}
Kernighan et al \cite{kernighan1970efficient} and Karypis et al  \cite{karypis1998multilevelk} develop techniques to partition the dataflow graphs which can be used to obtain model parallelism for training of deep neural networks.
Kerningham and Lin \cite{kernighan1970efficient} propose a formulation for the modeling of communication cost that can be used as the basis for dataflow graph partitioning for multi-device execution. Karypis and Kumar \cite{karypis1998multilevelk} develop a greedy refinement approach to partition the dataflow graph by first coarsening and then uncoarsening the graph.

The hardware schedulers have been explored mainly with the goal of reducing overheads in scheduling of jobs by an Operating System. We discuss some of the representative works. Eugen et al \cite{dodiu2012custom} design a hardware scheduler engine to reduce the task switching time targeted for Real Time Operating Systems (RTOSs). Gupta et al \cite{gupta2010hardware} devise a hardware scheduler to implement the {\textsf Pfair} scheduling algorithm which allows processes to make proportionate progress in a multi-processor system.

Our deep learning compiler performs partitioning of the dataflow graph like the other approaches mentioned above with some key differences: our techniques are applicable to directed graphs (and dataflow graphs are directed) whereas the Kerningham and Lin formulation is applicable only to undirected graphs. Unlike Karypis and Kumar’s approach where communication is the primary goal, and load balance is the secondary goal, in our formulation we can consider the load balance to be a primary goal as well which allows us to completely automatically achieve model parallelism.

The hardware scheduling assistant developed in this paper is intended to perform load balancing of work after being programmed by the deep learning compiler. In contrast, prior hardware schedulers are designed to assist scheduling of processes by an Operating System, which is a completely distinct problem.

\section{Conclusion}
\label{section:conclusion}
We presented a compiler technology and a hardware architecture to automatically achieve model parallelism during the training of deep neural networks. As the network size grows, the model will no longer fit in the memory of a single GPU or a single CPU. Therefore, it becomes imperative that the model parallelism be used to split the model across the memories of multiple compute devices while training. The compiler directed partitioning of the dataflow graph maps the computation to multiple compute devices and the hardware scheduling assistants dynamically adjust the mapping at runtime to maintain high load balance and low communication.

\balance
\bibliographystyle{ACM-Reference-Format}
\bibliography{paper}

\end{document}